# CellSecure: Securing Image Data in Industrial Internet-of-Things via Cellular Automata and Chaos-Based Encryption

Hassan Ali
*Department of Electrical Engineering*
*HITEC University Taxila*
Taxila, Pakistan
engr.hassan.ali.238@gmail.com

Muhammad Shahbaz Khan
*School of Computing, Engineering and the Built Environment*
*Edinburgh Napier University*
Edinburgh, UK
muhammadshahbaz.khan@napier.ac.uk

Maha Driss[1,2]
[1]*Robotics and Internet-of-Things Laboratory*
*Prince Sultan University*
Riyadh, Saudi Arabia
[2]*RIADI Laboratory, National School of Computer Sciences*
*University of Manouba*
Manouba, Tunisia
mdriss@psu.edu.sa

Jawad Ahmad
*School of Computing, Engineering and the Built Environment*
*Edinburgh Napier University*
Edinburgh, UK
j.ahmad@napier.ac.uk

William J. Buchanan
*School of Computing, Engineering and the Built Environment*
*Edinburgh Napier University*
Edinburgh, UK
b.buchanan@napier.ac.uk

Nikolaos Pitropakis
*School of Computing, Engineering and the Built Environment*
*Edinburgh Napier University*
Edinburgh, UK
n.pitropakis@napier.ac.uk

*Abstract*— In the era of Industrial IoT (IIoT) and Industry 4.0, ensuring secure data transmission has become a critical concern. Among other data types, images are widely transmitted and utilized across various IIoT applications, ranging from sensor-generated visual data and real-time remote monitoring to quality control in production lines. The encryption of these images is essential for maintaining operational integrity, data confidentiality, and seamless integration with analytics platforms. This paper addresses these critical concerns by proposing a robust image encryption algorithm tailored for IIoT and Cyber-Physical Systems (CPS). The algorithm combines Rule-30 cellular automata with chaotic scrambling and substitution. The Rule 30 cellular automata serves as an efficient mechanism for generating pseudo-random sequences that enable fast encryption and decryption cycles suitable for real-time sensor data in industrial settings. Most importantly, it induces non-linearity in the encryption algorithm. Furthermore, to increase the chaotic range and keyspace of the algorithm, which is vital for security in distributed industrial networks, a hybrid chaotic map, i.e., logistic-sine map is utilized. Extensive security analysis has been carried out to validate the efficacy of the proposed algorithm. Results indicate that our algorithm achieves close-to-ideal values, with an entropy of 7.99 and a correlation of 0.002. This enhances the algorithm's resilience against potential cyber-attacks in the industrial domain.

## I. INTRODUCTION

As Industry 4.0 and the Industrial Internet of Things (IIoT) have become integral components of modern industrial ecosystems, securing data transmission in industrial applications has become critically important. Among other data types, , images are widely transmitted and utilized across various IIoT applications, such as real-time remote monitoring, sensor-generated visual information, and quality assurance in production lines. To secure this image data not only efficient computational framework are required but there is also need of robust encryption algorithms capable of withstanding various forms of cyber threats [1]. Traditional encryption methods like AES [2] and DES [3] are not well suited for the complex demands of image encryption [4, 5]. This lead to a growing interest in exploring new techniques for fast and efficient encryption, such as chaos [6, 7]. Chaos-based encryption is preferred over traditional methods [8] due to its inherent characteristics like unpredictability, pseudo-randomness, high sensitivity to control parameters, and ergodicity [9]. Many encryption schemes employ simple one-dimensional chaotic maps [10, 11], but they have a limited chaotic range. In contrast, multi-dimensional chaotic maps [12] offer greater complexity but come at a higher computational cost. Therefore, hybrid chaotic maps have been introduced [13], balancing computational efficiency with a large chaotic region.

Moreover, there has been an interest in developing various cryptosystems incorporating a lightweight and efficient pseudo-random number generator (PRNG) called cellular automata. Cellular automata are discrete dynamical systems in both space and time [14]. They consist of an array of cells, with each cell capable of assuming a value from a limited set of possibilities. These cells are updated synchronously at discrete time intervals according to a specific interaction rule [15]. Recently, the Rule 30 [16] has been employed to produce numbers exhibiting high randomness. In this paper, we propose a robust image encryption algorithm, tailored for IIoT and Cyber-Physical Systems (CPS) in the era of Industry 4.0. The proposed algorithm is a combination of a hybrid chaotic system, i.e., Logistic Sine System (LSS) and Rule 30 of cellular automata. The scheme is divided in 3 stages: chaotic shuffling, chaotic substitution, and Rule 30 cellular automata.

Main contributions of this paper are:

1. The introduction of a robust image encryption algorithm tailored for IIoT and cyber-physical systems that integrates Rule-30 cellular automata with chaotic scrambling and substitution. This algorithm offers inherent unpredictability and complex patterns that induce non-linearity, which are crucial for enhanced security.

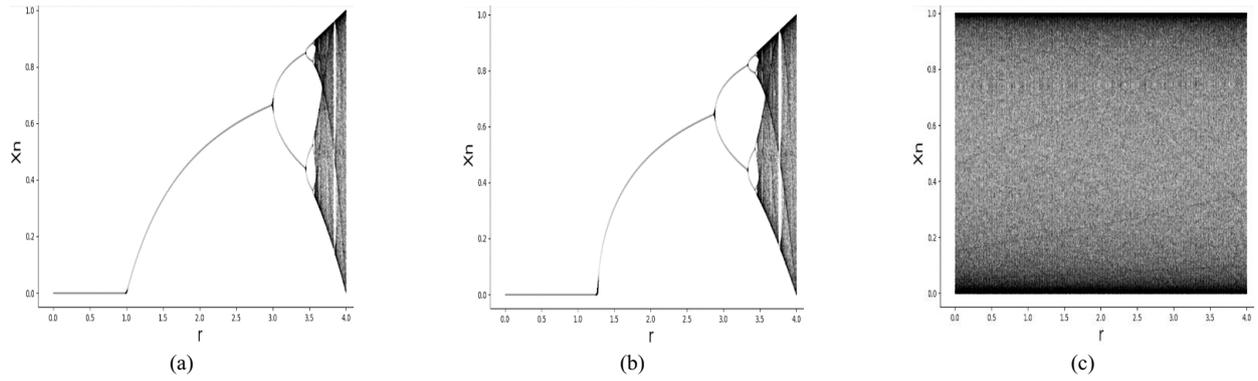

Fig. 1. Bifurcation Diagram of Chaotic Maps; (a) Logistic Map, (b) Sine Map, (c) Logistic Sine System.

2. This paper emphasizes the lightweight nature of the cellular automata approach, making the proposed algorithm highly suitable for real-time industrial applications requiring rapid encryption and decryption cycles.

3. An extensive security analysis validating the robustness of the proposed algorithm. The algorithm achieves close-to-ideal entropy and correlation values, validating its resilience against potential cyber-attacks in modern industrial systems.

## II. CELLSECURE: THE PROPOSED ENCRYPTION SCHEME

The proposed encryption scheme comprises multiple stage of confusion and diffusion driven by chaotic Logistic-Sine Map and Rule 30 Cellular Automata, and is discussed in detail below.

### A. Logistic-Sine Map

This paper utilizes a Logistic-Sine system [13], which is a combination of Logistic map (L) and Sine map (S) to provide a new hybrid chaotic map as defined in (1) and (2).

$$X_{N+1} = \left(L(r, X_N) + S\big((4-r), X_N\big)\right) \bmod 1 \quad (1)$$

$$X_{N+1} = \left(r X_N (1 - X_N) + \frac{(4-r)\sin(\pi X_N)}{4}\right) \bmod 1 \quad (2)$$

Where $L(r, X_N)$ represents the logistic map portion and $S\big((4-r), X_N\big)$ represents the sine map portion. The parameter $r \in (0, 4)$. Fig. 1 show the bifurcation diagram of Logistic map, Sine map and the combined hybrid LSS. It can be observed that the chaotic region of LSS is large as compared to the Logistic and Sine maps.

### B. Rule-30 Cellular Automata

The proposed algorithm utilizes an one-dimensional cellular automaton rule, i.e., Rule 30. Rule 30 operates on each cell and its two neighbors to form a neighborhood of 3 cells (left, center, right). Each cell is either in on state (1) or off state (0) and through these simple steps shown below we can see how each state of cell evolves based on its neighbors. Mathematically, Rule 30 gives the next state of any given cell as equation (3).

$$M_i(t+1) = M_{i-1}(t) \otimes \big(M_i(t) \vee M_{i+1}(t)\big) \quad (3)$$

Where $M_i(t)$ represents the state of the cell at position $i$ at time $t$. The symbols $\otimes$ and $\vee$ represent the XOR and OR Boolean operations. The transition rules are summarized in a truth table, given in Table 1. The first three columns represent the states of the left, center, and right cells, respectively, and the last column represents the next state of the center cell. The pattern of Rule 30 cellular automata for first 10 and 100 steps based on the transition rules are given in Fig.2 (a) and Fig. 2(b), respectively and first four rules are described as follows:

- If the left and center cells are in the off state (0) and the right cell is in the on state (1), the center cell in the next generation becomes on (1).

- If the left cell is in the off state (0) and the center and right cells are in the on state (1), the center cell in the next generation becomes on (1).

- If the left and center cells are in the on state (1) and the right cell is in the off state (0), the center cell in the next generation becomes on (1).

- If all three neighbors are in the off state (0), the center cell in the next generation remains off (0).

### C. The Proposed Encryption Algorithm

The proposed encryption process given in Fig 3 proposes a new algorithm combining a hybrid chaotic maps based shuffling and substitution, and Rule 30 cellular automata to create a robust and secure scheme.

#### 1) Chaotic Shuffling

Firstly, we generate two arrays containing random values equal to the rows and columns of the image using the LSS. These arrays are sorted to find random indices through which we shuffle rows and columns of the plaintext image to get a shuffled image (**S2**). The shuffling algorithm is given as Algorithm 1.

Table I.  TRANSITION RULES OF RULE 30

| $M_{i-1}(t)$ | $M_i(t)$ | $M_{i+1}(t)$ | $M_i(t+1)$ |
|---|---|---|---|
| 0 | 0 | 1 | 1 |
| 0 | 1 | 1 | 1 |
| 1 | 1 | 0 | 1 |
| 0 | 0 | 0 | 0 |
| 1 | 0 | 0 | 0 |
| 0 | 1 | 0 | 0 |
| 1 | 0 | 1 | 0 |
| 1 | 1 | 1 | 0 |

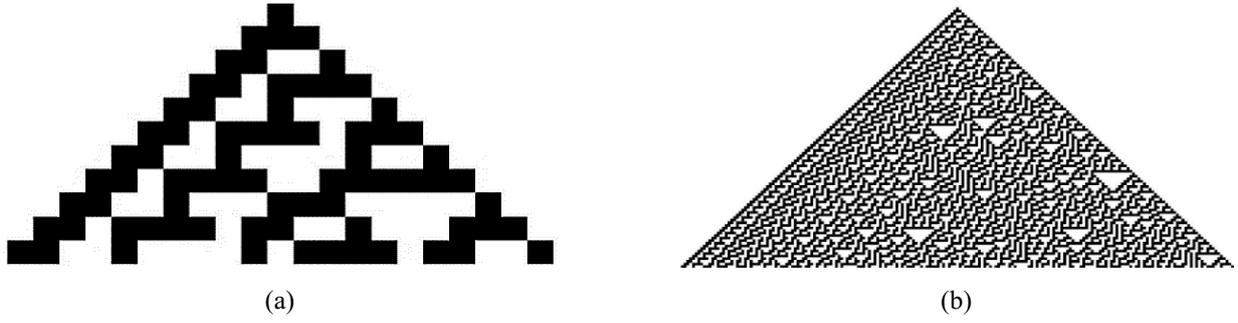

(a)                  (b)

Fig. 2. Patten for the: (a) First 10 Steps of Rule 30, (b) First 100 Steps of Rule 30 [15]

*3) Rule 30 Cellular Automata*

Lastly, the LSS is used to generate a random 16x16 matrix (*M*) containing values between 0 and 255 which is then further shuffled using the Rule 30 of the Cellular Automata to further create a more complex and random matrix by considering the right left and center value of the pixels on which XOR and OR operations are performed. Each pixel is changed and this process is repeated for an iteration (*I*) times to achieve even more randomness as represented in Fig. 4. The pseudo code algorithm for cellular automata is given in Algorithm 3 and the following steps entail how cellular automata is applied.

- **Initialization:** Let the initial state of the cellular automaton with n cells is the current state.

$$curr\_St = [M_1(0), M_2(0), \cdots, M_n(0)] \quad (4)$$

- **Evolution:** For $t = 1,2,\cdots,T$, where $T$ is the number of steps following steps were implemented:

  1) An array of length n for the next sate was initialized.

  $$nextState = [0,0,\cdots,0] \quad (5)$$

  2) For $i = 2$ to $i = n - 1$, the $nextState[i]$ was updated using the Rule 30 equation.

  $$nextState[i] = curr\_St[i-1] \oplus \binom{curr\_St[i] \lor}{curr\_St[i+1]} \quad (6)$$

  3) Finally we set, $curr\_St = nextState$.

*4) Ciphertext Image*

After step 4, a new matrix is achieved which is then finally used to bitwise-XOR each 16x16 block of the substituted image to achieve our final ciphertext image (*C*).

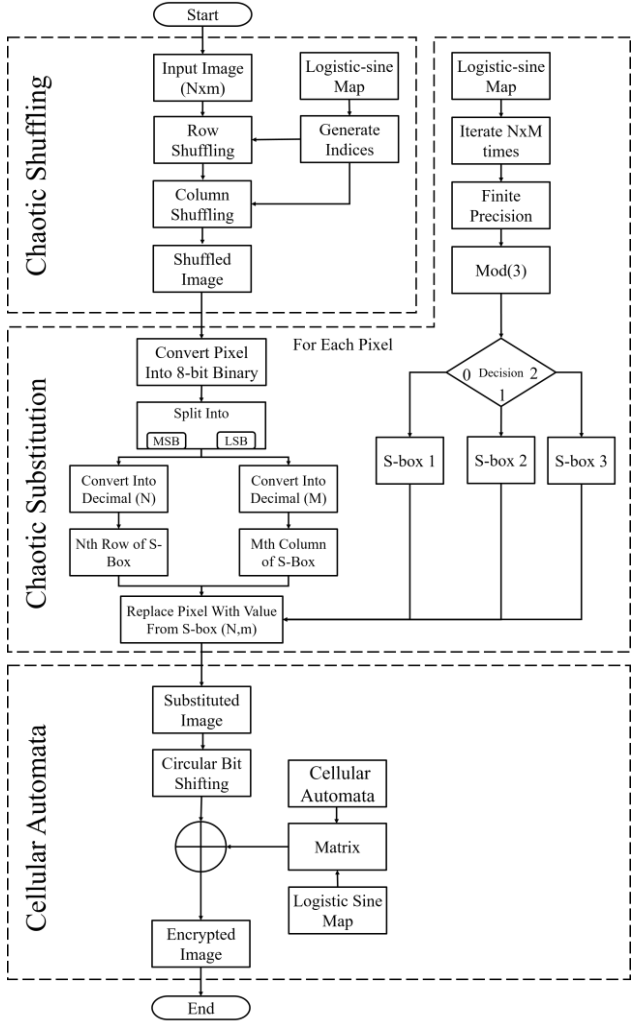

Fig. 3. The Proposed Encryption Algorithm

*2) Chaotic Substitution*

Then, we generate a 256×256 matrix using LSS. We use finite precision and modulo 3 operation to generate a random array that had only 0, 1, and 2 as values. This array helped us select a specific S-Box to change the pixel value of the shuffled image. Each pixel was converted into 8-bit binary, which we further split into two 4-bit parts (MSB, LSB), and changed them back into decimal numbers to obtain an index values. We used these indices to pick values from the chosen S-Box and change the pixel values. This gave us a new image (*Ŝ*). The pseudo code for chaotic substitution is given in Algorithm 2.

| ALGORITHM 1: | CHAOTIC SHUFFLING |
|---|---|
| | **Inputs:** Plaintext Image (*P*), Initial Keys (r, x₀) |
| | **Outputs:** shuffled image (*S2*) |
| 1. | [H, W] = size(*P*) |
| 2. | **Function** shuffle(H, W, r, x₀) |
| 3. | **For** i = 1 : H |
| 4. |    X[i-1]= x₀ |
| 5. |    X[i] = LSS(X[i-1], r) |
| 6. | **End For** |
| 7. | **For** i = 1 : W |
| 8. |    Y[i-1] = x₀ |
| 9. |    Y[i] = LSS(Y[i-1], r) |
| 10. | **End For** |
| 11 | row_indices, col_indices = sort(X), sort(Y) |
| 12 | *S1*= *P* [row_indices,:] |
| 13 | *S2*= *S1* [:, col_indices] |
| 14 | **Return** *S2* |
| 15 | **End Function** |

**ALGORITHM 2: CHAOTIC SUBSTITUTION**
**Inputs:** Shuffled Image ($S$), Initial Keys ($r$, $x_o$)
**Outputs:** Substituted Image ($\hat{S}$)
1. [H, W] = size($S2$)
2. **For** i = 1 : HxW
3.     V[i-1] = $x_o$
4.     V[i] = LSS(V[i-1], r)
5. **End For**
6. V1 = round(Vx10$^3$)
7. sequence_sbox = MOD(V1, 3)
8. **Function** substitution(value)
9.     binary_input = binary(value, 8-bit)
10.     msb_input = binary_input[:4]
11.     lsb_input = binary_input[4:]
12.     Msb$_{dec}$ = integer(msb_input, 2)
13.     Lsb$_{dec}$ = integer(lsb_input, 2)
14.     Return Msb$_{dec}$, Lsb$_{dec}$
15. **End Function**
16. K = 0
17. **For** i = 1 : H
18.    **For** j = 1 : W
19.     If sequence$_{sbox}$[K] = 0
20.      S-box = S-box1
21.     Else If sequence$_{sbox}$[K] = 1
22.      S-box = S-box2
23.     Else If sequence$_{sbox}$[K] = 2
24.      S-box = S-box3
25.     **End If**
26.     value = $S2$[i, j]
27.     Msb$_{dec}$, Lsb$_{dec}$ = substitution(value)
28.     output = S-box[Msb$_{dec}$, Lsb$_{dec}$]
29.     $\hat{S}$[i, j] = output
30.     K = K+1
31.    **End For**
32. **End For**

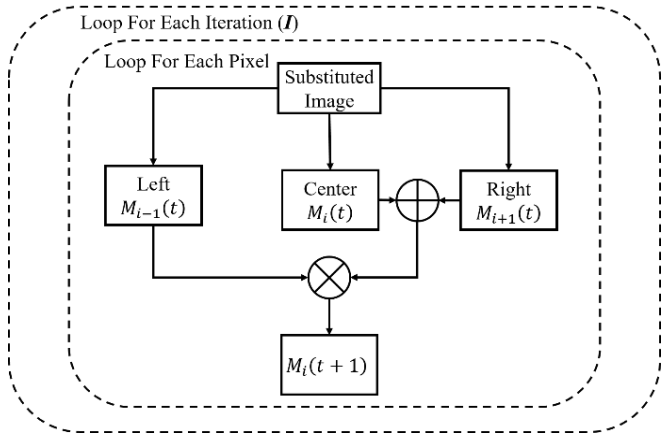

Fig. 4. Visual Representation of Rule 30

## III. RESULTS AND ANALYSIS

This section outlines the results of our proposed encryption scheme. Two gray test images (Baboon and Cameraman) of size 256x256 were used and extensive security analysis was carried out.

### A. Histogram Analysis

Histogram analysis refers to the study of the distribution of pixel values in an image. In the context of image encryption, a good encryption algorithm should produce an encrypted image with a histogram that is substantially different from the original, making it hard to glean any information about the original image's content. Ideally, the histogram of the encrypted image should be flat, meaning that all pixel values are equally probable, ensuring maximum randomness and minimizing the chances of decryption without the correct key. It can be seen in Fig. 5 that histograms of the encrypted images are evenly distributed, showing effectiveness of the proposed scheme.

### B. Statistical Security Analysis

The statistical security parameters, i.e., entropy, correlation, energy, contrast and homogeneity have been calculated. Table 2 presents the results of the statistical security analysis. The results prove the effectiveness of our scheme, with close to ideal values, i.e., an entropy 7.99 for both Cameraman and Baboon images, homogeneity of 0.388 for both images, a close to 0 energy of 0.0156 for both images, and a contrast as large as possible of 10.6411 and 10.5361 for the Baboon and Cameraman, respectively. Moreover, Fig. 5 provides the histogram analysis of the plaintext and cipher images.

*1) Entropy*

In image encryption, entropy measures the randomness or unpredictability of pixel values in an encrypted image. A higher entropy value typically indicates a more secure encryption, as the data appears more random and harder to decode. Ideally, for an 8-bit grayscale image, the maximum entropy value is 8, signifying a perfectly random distribution of pixel values. Evaluating entropy helps in determining the strength and effectiveness of an encryption algorithm: the closer the entropy is to its maximum value, the more secure the encryption. It can be seen from Table 2 that entropy of both test images are approximately 8.

*2) Homogeneity*

Homogeneity refers to the consistency or uniformity of pixel values in an encrypted image. In a well-encrypted image, pixel values are distributed uniformly, making patterns hard to discern. Therefore, a high degree of homogeneity suggests that an encryption algorithm is successful in obscuring any noticeable patterns or structures, further ensuring the encrypted image is secure and resistant to various attacks. Evaluating homogeneity assists in gauging the quality of encryption and its potential vulnerability to pattern-based decryption attempts. Homogeneity of both test images are given in Table 2.

**ALGORITHM 3: CELLULAR AUTOMATA**
**Inputs:** Substituted Image ($\hat{S}$), Initial Keys ($r$, $x_o$), Iteration ($I$)
**Outputs:** Ciphertext Image ($C$)
1. **For** I = 1 : 256
2.    Z[i-1] = $x_o$
3.    Z[i] = LSS(Z[i-1], r)
4. **End For**
5. matrix = reshape(sort(Z), (16x16))
6. **Function** apply_rule30(left, center, right):
7.    return left ^ (center | right)
8. **End Function**
9. **Function** rule30(value)
10.    **For** _ in range($I$)
11.     **For** i in range(1, H - 1):
12.      **For** j in range(W):
13.       left = encrypted_matrix[i - 1, j] if i > 0 else 0
14.       center = encrypted_matrix[i, j]
15.       right = encrypted_matrix[i + 1, j] if i < H - 1 else 0
16.       temp_matrix[i, j] = apply_rule30(left, center, right)
17.    encrypted_matrix = np.copy(temp_matrix)
18.    return encrypted_matrix
19. **End Function**
20. matrix = rule30(matrix)
21. Bit-matrix = repmat(matrix, H/16, W/16)
22. $C$ = bitwise-XOR($\hat{S}$, Bit-matrix)

*3) Energy*

Energy refers to the intensity of pixel values in an encrypted image. When an image is encrypted effectively, its energy should spread out, making the image appear as a mix of intensities. A low energy value indicates that the encryption has disrupted the original image's features, making it difficult for unauthorized users to extract information. Energy of the test images in given in Table 2.

*4) Contrast*

Contrast is the difference in color due to which objects appear different in an image. A successful encryption process often transforms the image's contrast, making the details harder to distinguish and the image difficult to interpret without the decryption key. The contrast should be high, which can be seen in the results given in Table 2.

Table II. STATISTICAL TEST ANALYSIS

| Images | Entropy | Homogeneity | Contrast | Energy |
|---|---|---|---|---|
| Baboon | 7.9958 | 0.3889 | 10.6411 | 0.0156 |
| Cameraman | 7.9944 | 0.3888 | 10.5361 | 0.0156 |

Table III. CORRELATION COEFFICIENTS OF ADJACENT PIXELS OF CIPHERTEXT IMAGE

| Images | Vertical | Horizontal | Diagonal |
|---|---|---|---|
| Baboon | -0.0048808 | -0.0010709 | 0.0047433 |
| Cameraman | 0.0022096 | -0.00446 | 0.0043268 |

Table IV. CORRELATION COEFFICIENTS BETWEEN PLAINTEXT AND CIPHERTEXT IMAGE

| Images | Vertical | Horizontal | Diagonal |
|---|---|---|---|
| Baboon | -0.010943 | -0.003131 | -0.0086117 |
| Cameraman | -0.0021706 | -0.0032932 | -0.0029474 |

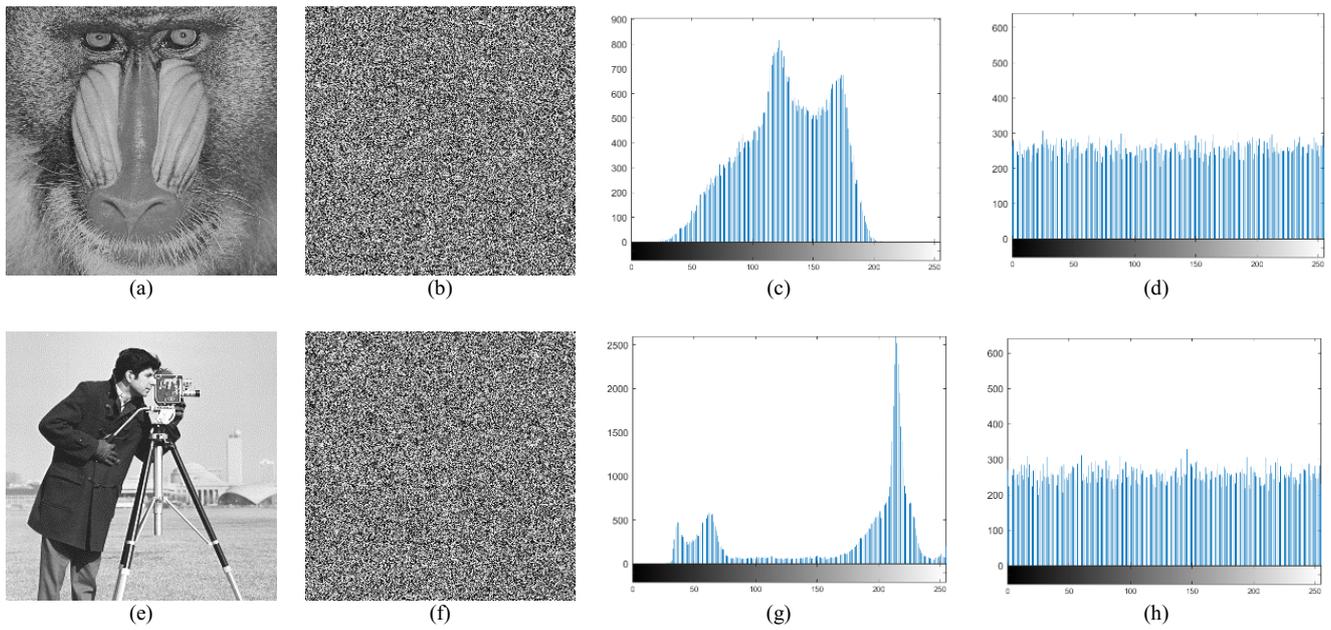

Fig. 5. Histogram Analysis of Plaintext and Ciphertext Images. Baboon(a)-(d), Cameraman(e)-(h)

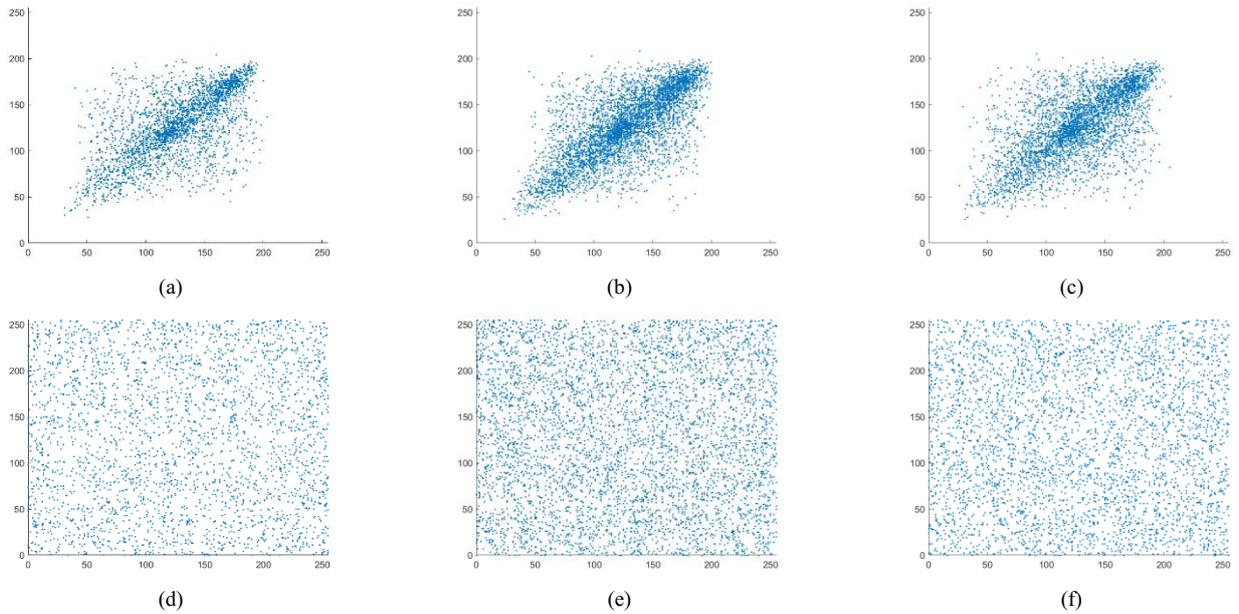

Fig. 6. Correlation Analysis of Baboon Image: Vertical, Horizontal and Diagonal Coefficients of Plaintext Image(a)-(c); Vertical, Horizontal and Diagonal Coefficients of Ciphertext Image(d)-(f)

*C. Correlation Analysis*

Correlation refers to the statistical relationship between adjacent pixel values in an image matrix. The objective is to minimize this correlation in the encrypted image so that adjacent pixels appear to be independent. If two encrypted images have a correlation value close to 0, they don't share obvious patterns and are considered different from each other. Mathematically correlation coefficient cab be represented as follows:

$$CorrC = \frac{\sum_{i=1}^{M}\sum_{j=1}^{N}\begin{pmatrix}P(i,j)-E(P)\\C(i,j)-E(C)\end{pmatrix}}{\sqrt{\sum_{i=1}^{M}\sum_{j=1}^{N}(P(i,j)-E(P))^2 \sum_{i=1}^{H}\sum_{j=1}^{N}(C(i,j)-E(C))^2}} \quad (7)$$

Where,

- $\sum_{i=1}^{M}\sum_{j=1}^{N}(\cdots)$ and $\sum_{i=1}^{H}\sum_{j=1}^{N}(\cdots)$ are summations over the pixels of the images, presumably for all rows $i$ and all columns $j$.
- $P(i,j)$ and $C(i,j)$ are the pixel values at specific location.
- $E(P)$ and $E(C)$ are the expected values.

The correlation of the test images and its cipher is visually displayed in Fig. 6. It can be seen that the correlations in the ciphertext image are broken successfully. Table 3 provides the correlation between adjacent pixels of the ciphertext and Table 4 provides correlation between plaintext and ciphertext. All vertical, horizontal, and diagonal coefficients are closer to zero which shows that the strong correlation of the image is broken.

## IV. CONCLUSION

Given the emergent challenges posed by the Industrial IoT (IIoT) and Industry 4.0 landscapes, this paper successfully introduced a robust image encryption algorithm specifically designed to meet the security needs of IIoT and Cyber-Physical Systems (CPS). The proposed algorithm combines of Rule-30 cellular automata and a hybrid chaotic map, the logistic-sine map, to produce an encryption framework that is both computationally efficient and secure. Performance evaluations confirm the algorithm's high resilience against statistical attacks, achieving near-optimal entropy and correlation values of 7.99 and 0.002, respectively. As a future work, we plan to further enhance the algorithm's security by incorporating secure hash algorithms (SHA) for key generation, adding an additional layer of protection against differential attacks. This work lays a solid foundation for the secure handling of image data in the industrial domain.

## ACKNOWLEDGMENT

The research leading to these results has been partially supported by the Horizon Europe Project Trust & Privacy Preserving Computing Platform for Cross-Border Federation of Data (TRUSTEE), (GA 101070214).